\documentclass[a4paper]{jpconf}
\usepackage{graphicx}
\usepackage{subfigure}
\begin{document}
\title{Results from the NEXT prototypes}

\author{C A B Oliveira, on behalf of the NEXT collaboration}

\address{Lawrence Berkeley National Laboratory\\
One Cyclotron Road, Berkeley, CA 94720, USA}

\ead{cabdoliveira@lbl.gov}

\begin{abstract}
NEXT-100 is an electroluminescent high pressure Time Projection Chamber currently under construction. It will search for the neutrino-less double beta decay in $^{136}$Xe at the Canfranc Underground Laboratory. NEXT-100 aims to achieve nearly intrinsic energy resolution and to highly suppress background events by taking advantage  of the unique properties of xenon in the gaseous phase as the detection medium. In order to prove the principle of operation and to study which are the best operational conditions, two prototypes were constructed: NEXT-DEMO and NEXT-DBDM. In this paper we present the latest results from both prototypes. We report the improvement in terms of light collection ($\sim3\times$) achieved by coating the walls of NEXT-DEMO with tetraphenyl butadiene (TPB), the outstanding energy resolution of 1 \% (Full Width Half Maximum) from NEXT-DBDM as well as the tracking capabilities of this prototype (2.1 mm RMS error for point-like depositions) achieved by using a square array of $8\times8$ SiPMs. 
\end{abstract}

\section{Introduction}

If the neutrino has a Majorana nature, the neutrino less double beta ($0\nu\beta\beta$) decay exists in some natural isotopes including $^{136}$Xe. The observation of this hypothetical decay would prove physics beyond the Standard Model as the neutrino being its own anti-particle and the lepton number violation. In addition, the measurement of its lifetime would allow to estimate the absolute neutrino masses and to determine the hierarchy of the neutrino mass states.

NEXT-100 is currently under construction and will search for the $0\nu\beta\beta$ decay at the Canfranc Underground Laboratory~\cite{LSC}. It consists of an electroluminescent high pressure (up to 15 bar) Time Projection Chamber that will use up to 150 kg of gaseous xenon enriched in the isotope $^{136}$Xe to $\sim91$ \% as both the source of the decay and the detection medium.

In the gaseous phase the fluctuations associated with the production of primary electrons by ionizing particles are very low when compared to those of the liquid phase ($\sim$130 times higher)~\cite{bolotnikov}. Taking advantage of this and using electroluminescence (EL) as a nearly noiseless signal amplification process~\cite{EL_coliveira} allows one to achieve outstanding energy resolution, which plays an important role in separating the rare signal from natural background decays that are close in energy. In addition, the extended ionization tracks left behind by the two high energy electrons from the decay ($\sim30$ cm long at 10 bar) enable the suppress of background by performing topology recognition of the ``spaghetti with two meat-balls'' signature~\cite{CDR}. For further details about NEXT-100, its working principle and its physics potential please see references~\cite{TDR}and~\cite{physics_potential}.

As part of the R\&D program of the NEXT experiment, two prototypes were constructed: NEXT-DEMO and NEXT-DBDM. Both prototypes have a similar working principle (and similar to that of NEXT-100) but the studies performed in each one focused on different issues. Unlike NEXT-100, the two prototypes are not intended to be radiopure neither shielded from natural radioactivity. In this paper we summarize the latest results from both prototypes regarding gas purity stability, light collection, energy resolution and tracking capabilities. Detailed analyses are available in references~\cite{NEXTDEMO_initial} and~\cite{NEXTDBDM_initial} for NEXT-DEMO and NEXT-DBDM, respectively.

\section{Principle of operation and data acquisition}

\begin{figure}[h]
\centering
\includegraphics[width=0.5\textwidth]{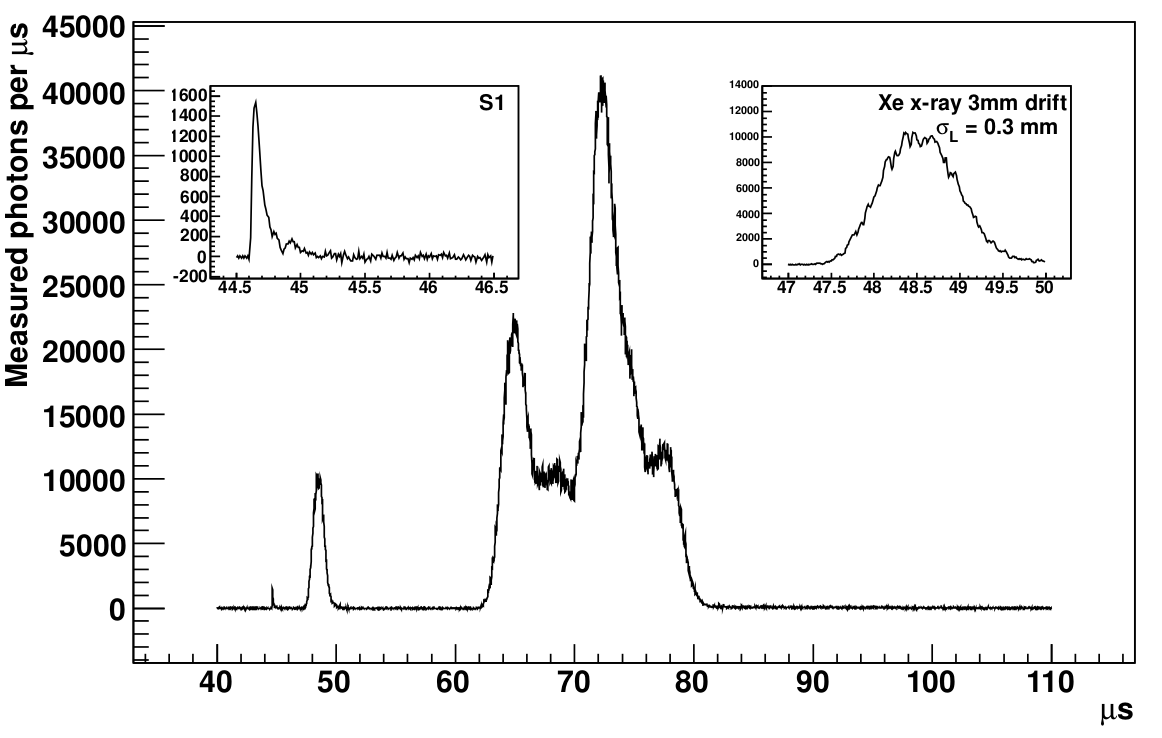}%
\caption{Sum of the waveforms recorded by the 19 PMTs of NEXT-DBDM for a 662 keV Gamma-ray interaction. At $\sim$ 45 $\mu$s it is visible the prompt S$_1$ pulse, at $\sim$ 48 $\mu$s a 30 keV xenon X-ray deposition pulse, for $t>65\mu$s the main event deposition which structure may be affected by the topology of the track, the formation of $\delta$-rays, Compton depositions and multiple scattering dependence on the electron energy.}
\label{fig:NEXTDBDM_waveform}
\end{figure}

When incoming particles interact inside the detector, they ionize and excite atoms of the gas. Excitations produce a prompt and fast light (172 nm) pulse -- called S$_1$ -- that, after being detected by an array of photomultipliers (PMTs), works as a \textit{start-of-event} signal. The free electrons are then under effect of a drift field that helps them to escape recombination while they drift towards a region -- the EL gap -- where they are accelerated by a stronger electric field and excite but not ionize xenon atoms, which then de-excite producing secondary light (also 172 nm). This process is called \textit{electroluminescence} (EL) and, being detected by the array of PMTs, this light gives information about the energy of the incident particle -- the S$_2$ signal. EL is also detected by a 2D array of light sensitive detectors (PMTs in the case of NEXT-DEMO and SiPMs in the case of NEXT-DBDM) that lives just next to the EL gap. By time slicing the waveforms coming from those devices, it is possible to image in 3D the original ionization track. As an example of the acquired raw data, Fig.~\ref{fig:NEXTDBDM_waveform} shows the so-called \textit{summed waveform} which consists of the sum of the waveforms recorded by the 19 PMTs of NEXT-DBDM for a given event (662 keV Gamma-ray deposition).

The working principle of the prototypes is very similar to that of the final NEXT-100 detector~\cite{TDR}. For further details about the EL process please see reference~\cite{EL_coliveira}.

\section{NEXT-DEMO}
\begin{figure}[ht]
\centering
\subfigure[Schematic of the NEXT-DEMO prototype showing the pressure vessel, the energy and coarse tracking PMT planes, the field cage and the feed-throughs.]{
   \includegraphics[height=0.33\textwidth] {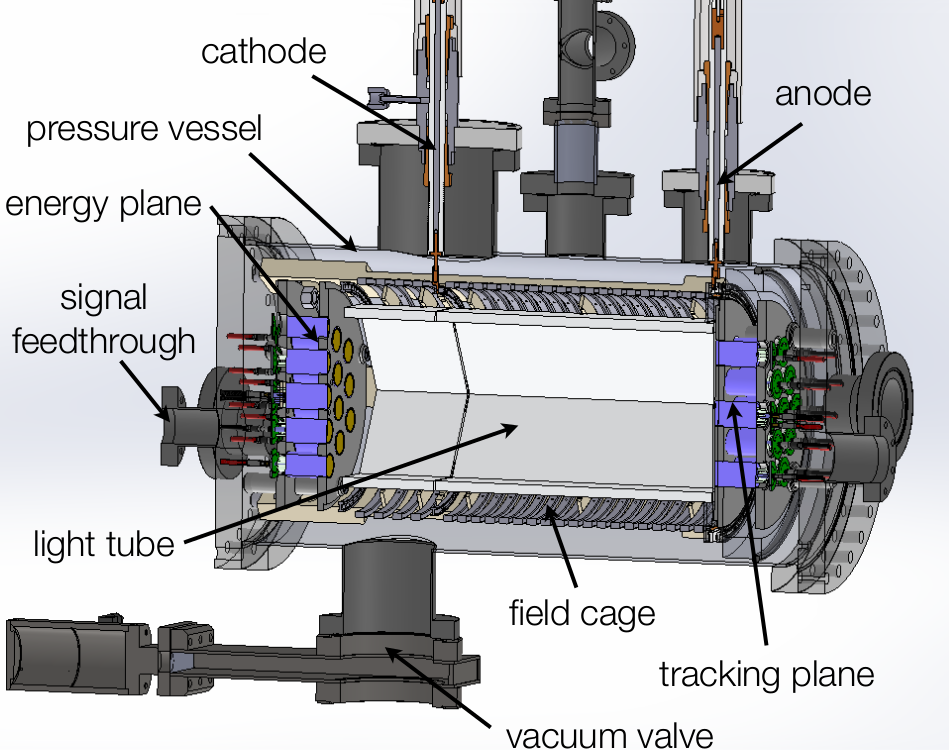}
   \label{fig:NEXTDEMO_schem}
 }
 ~
 \subfigure[Picture of the system showing the geometry of the $^{22}$ Na source and the NaI tagger setup relative to the pressure vessel.]{
   \includegraphics[height=0.33\textwidth] {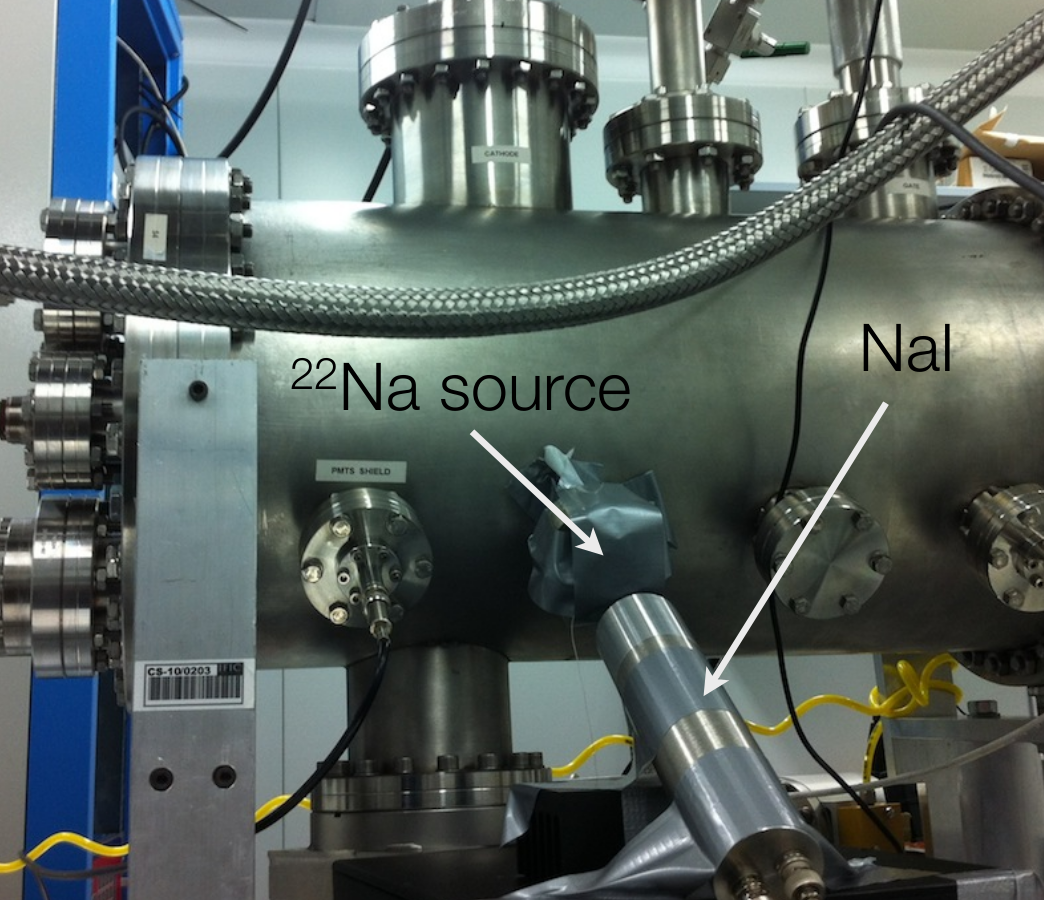}
   \label{fig:NEXTDEMO_source_pic}
 }
 \caption{NEXT-DEMO prototype setup.}
\label{fig:NEXTDEMO}
\end{figure}

NEXT-DEMO was constructed and operated at the Instituto de F\'{i}sica Corpuscular (Valencia, Spain). The energy plane consists of an array of 19 R7378A PMTs from Hamamatsu with 1'' quartz windows, resistant to pressure up to 20 bar and with a quantum efficiency (QE) of about 20 \%. It is placed 10 cm behind the cathode. A similar array was installed 2 mm away from the anode for coarse tracking. The drift region is 30 cm long and has an hexagonal cross-section with a 14.4 cm long internal diagonal. The length to diameter ratio of the EL light detected by the energy plane is $l/d = 40.0 \textrm{ cm} / 14.4 \textrm{ cm} = 2.8$. This ratio does not optimise light collection but allowed for the construction of a longer pressure vessel with flexibility for possible upgrades. A 1 $\mu$Ci $^{22}$Na calibration source has been used to characterise the detector. $^{22}$Na decays emitting a positron that travels less than 1 mm before annihilating with an electron and emitting two back-to-back 511 keV Gamma-rays. A NaI detector was used as a trigger in order to optimize the fraction of useful data. The setup is in such a way that a large fraction of the Gammas enter the chamber perpendicular to the longitudinal direction, close to the cathode.
The prototype scheme is shown in Fig.~\ref{fig:NEXTDEMO_schem} and a picture of the system and the radioactive source + NaI detector setup is shown in Fig.~\ref{fig:NEXTDEMO_source_pic}.

Two different sets of data were taken with this prototype: the first with the Teflon walls uncoated -- the \textit{UV configuration} -- and a second one with those walls coated with tetraphenyl butadiene (TPB). TPB shifts 172 nm S$_1$ and S$_2$ light  to the blue range ($\sim 430$ nm where the PMTs have higher QE), thus we call this setup the \textit{Blue configuration}. Xenon was recirculated at 10 bar and room temperature through a hot getter (SASE PS4-MT15) to remove impurities. High gas purity was achieved, as confirmed by the smallness of the required corrections for attachment to the measured energy, in the range 1 - 3 \% for the whole 30 cm long drift region, depending on whether the walls of the detector were coated or uncoated. A drift field of $\sim667$ Vcm$^{-1}$ and an EL field of 2.4 kVcm$^{-1}$bar$^{-1}$ were used for both sets of data.

\subsection{Light collection improvement by TPB}

\begin{figure}[h]
\centering
\includegraphics[width=.9\textwidth]{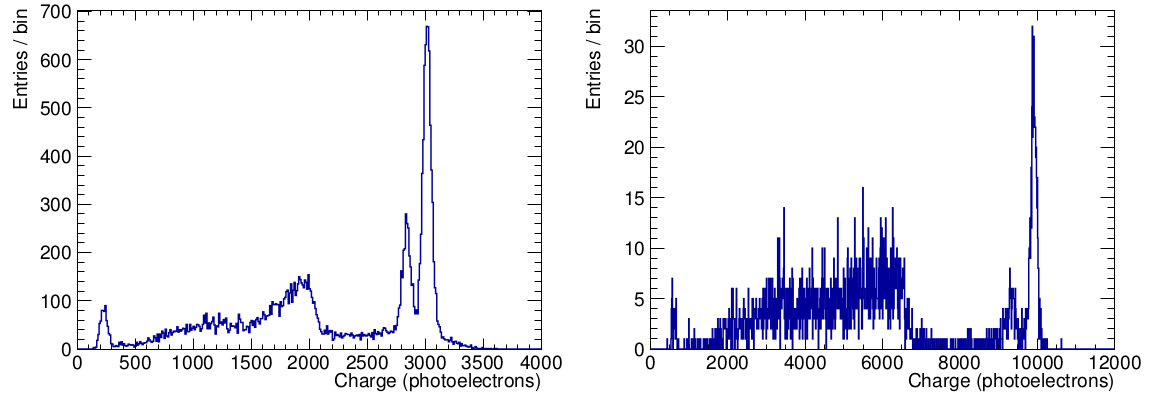}%
\caption{ Spectra obtained in NEXT-DEMO with the UV (left) and Blue (right) configurations, after applying corrections for attachment as well as for radial and angular position dependencies.}
\label{fig:NEXTDEMO_en_spectrum}
\end{figure}

Figure~\ref{fig:NEXTDEMO_en_spectrum} shows, in the left, the spectrum obtained with the UV configuration, after corrections for attachment as well as for radial and angular position dependencies, $\left(r,\phi\right)$, at the level of 0.5~-~3.0 \%. Only events with a reconstructed radial position less than 35 mm, where the position response of the detector is linear, are considered. From the left to the right of the $x$-axis, expressed in number of detected VUV photons - $pes$: at $\sim250$ $pes$ the 30 keV xenon fluorescence X-ray peak, between 500 and 2700 $pes$ the Compton shoulder, for $\sim2800$ $pes$ the escape peak and for $\sim3000$ $pes$ the full energy deposition peak. The corresponding spectrum obtained with the Blue configuration is shown in the right of Fig.~\ref{fig:NEXTDEMO_en_spectrum} and shows equivalent features, despite the lower statistics. The number of detected photons increases by a factor of $\sim3\times$ showing the advantage of coating the detector walls with TPB. The energy resolution of the full energy deposition peak (511 keV) is 2.8 \% FWHM for the UV configuration, in agreement with Monte Carlo simulations where only geometric effects were considered. For the Blue configuration, 1.7~\% FWHM is achieved, in agreement with the expected scaling due to the increase in the photon statistics. This shows that after calibration, the contributions to the energy response of the detector are well understood.

\section{NEXT-DBDM}

\begin{figure}[ht]
\centering
\subfigure[Schematic of the NEXT-DBDM prototype showing the pressure vessel, the energy plane, the field cage and the cathode and EL meshes.]{
   \includegraphics[height=0.28\textwidth] {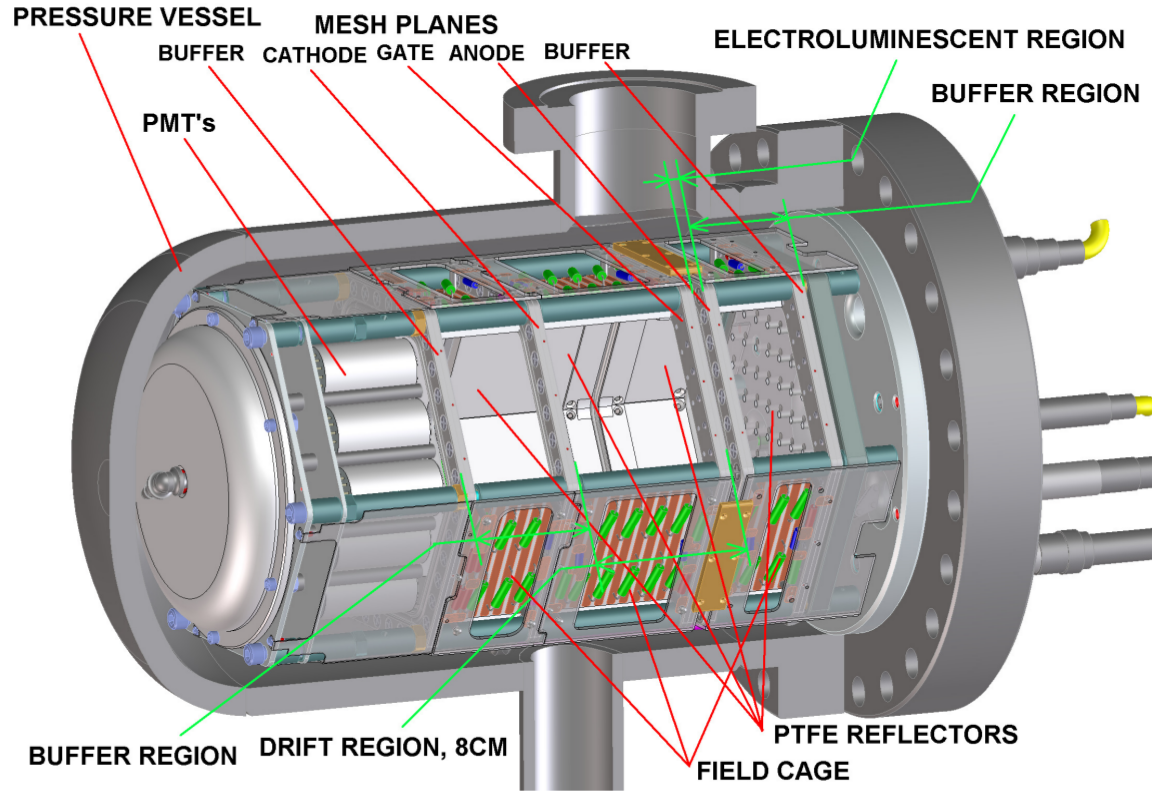}
   \label{fig:NEXTDBDM_Schem}
 }
 ~
 \subfigure[Picture of the system showing the pressure vessel, the feed-throughs that service the detector and the vaccum system.]{
   \includegraphics[height=0.28\textwidth] {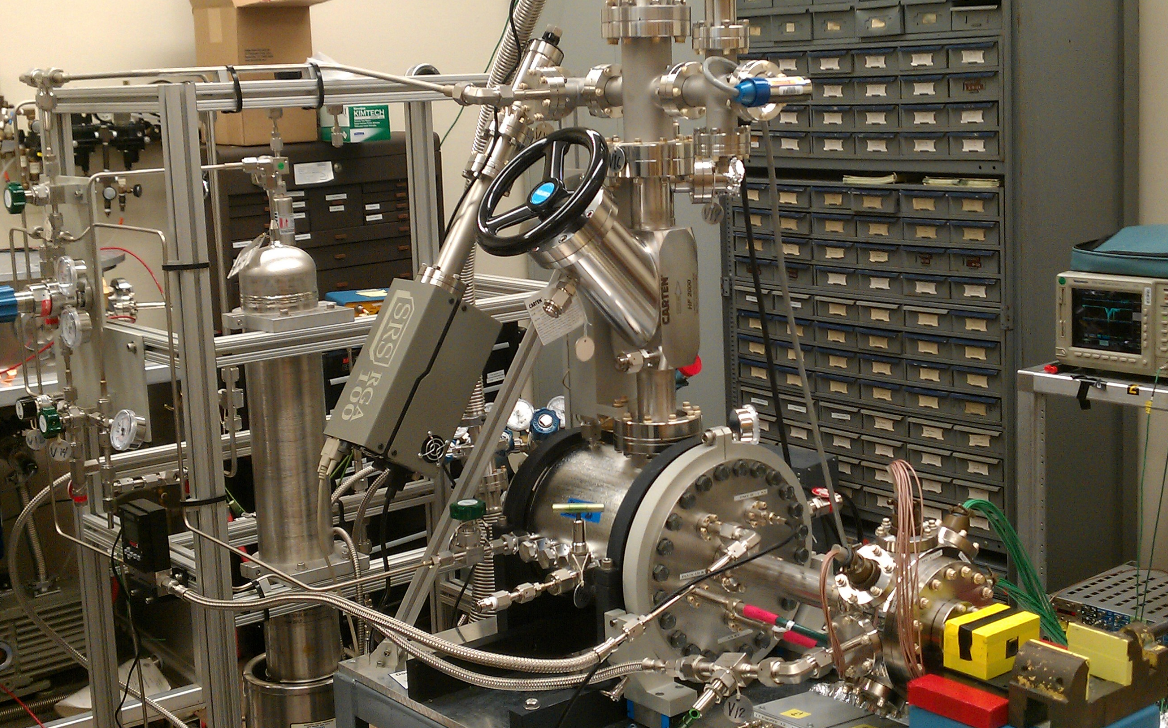}
   \label{fig:NEXTDBDM_pic}
 }
\label{fig:NEXTDBDM}
\caption{NEXT-DBDM prototype.}
\end{figure}

NEXT-DBDM is a prototype that aims to go beyond the NEXT R\&D and explore the possibility of Double Beta and Dark Matter (DBDM) searches using the same detector. It was constructed and is being operated in the Lawrence Berkeley National Laboratory (Berkeley, California, USA). Figure~\ref{fig:NEXTDBDM_Schem} shows the schematics of the detector; a picture of the pressure vessel, feed-throughs and vacuum system  is shown in Fig.~\ref{fig:NEXTDBDM_pic}.

The energy plane is made of 19 PMTs (same as those used in NEXT-DEMO) placed 5 cm behind the cathode. The drift region is 8 cm long with an hexagonal cross section of 14.4 cm diameter. The length to diameter associated with the EL light is $l/d = 13.0/14.4\mathrm{ cm} = 0.9$, optimizing light collection. This allowed us to perform detailed energy response studies envisaging nearly intrinsic resolutions. The results reported in this document refer to events produced by a 1 mCi $^{137}$Cs 662 keV Gamma-ray source, after being strongly collimated and directed along the axis of the chamber.

Xenon at room temperate and pressures of 10 and 15 bar was used, recirculating through a hot getter from Johnson Matthey (model PureGuard) for purification. High quality pure xenon was achieved, indicated by the long electronic life times: $36\pm10$ ms for 10 bar and $9.1\pm0.9$ ms for 15 bar. This means that less than 1 \% of the primary electrons were lost through attachment along the whole 8 cm long drift region. 

Drift fields in the range of 10 - 120 Vcm$^{-1}$bar$^{-1}$ were scanned without significant changes in the obtained energy resolution. EL amplification fields between 1.0 and 2.5 kVcm$^{-1}$bar$^{-1}$ were also scanned and good consistency of the amount of EL light dependency on the field was found for both pressures.

\subsection{Energy studies}
\label{subsec:re}

Results reported in this Section correspond to runs taken without a dedicated tracking plane. However, a rough $\left(x,y\right)$ reconstructed position of the events was possible using the energy PMT plane, with resolutions of $\sim2$ mm for central 30 keV X-ray depositions.

\begin{figure}[ht]
\centering
\subfigure[Energy resolution for 662 keV depositions at 10.1~atm. The data was acquired with a 0.16 kVcm$^{-1}$ drift field and a EL field of 2.08 kVcm$^{-1}$atm$^{-1}$. The events were corrected for small attachment losses corresponding to a life time of $\tau=13.9$ ms.]{
   \includegraphics[width=0.44\textwidth] {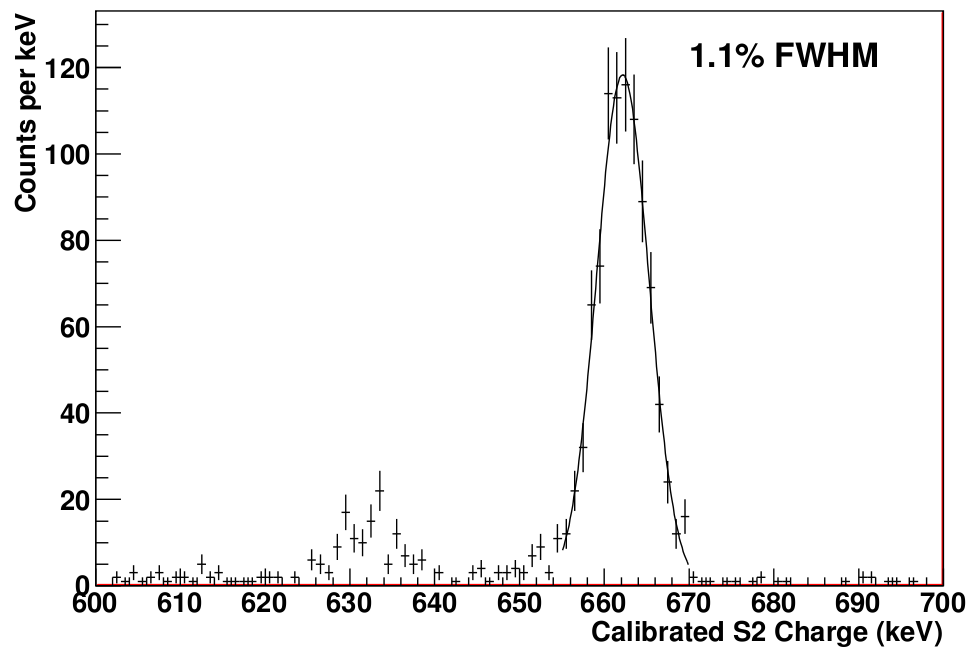}
   \label{fig:NEXTDBDM_EnRes_662_10bar}
 }
 ~
 \subfigure[Energy spectrum for fluorescence xenon X-rays acquired at 10.1 bar. The data was acquired using a drift field of 1.03 kVcm$^{-1}$ and an EL field of 2.68 kVcm$^{-1}$atm$^{-1}$.]{
   \includegraphics[width=0.44\textwidth] {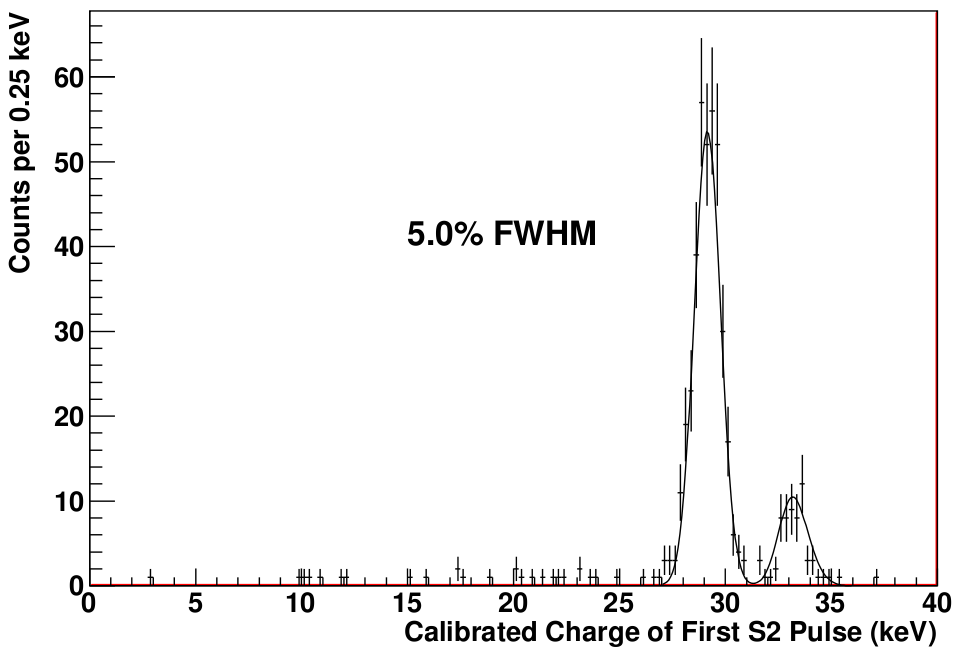}
   \label{fig:NEXTDBDM_EnRes_30_10bar}
 }
\label{fig:NEXTDBDM_EnRes}
\caption{Energy resolution with NEXT-DBDM.}
\end{figure}

Figure~\ref{fig:NEXTDBDM_EnRes_662_10bar} shows the acquired energy spectrum at 10 atm around the 662 keV energy window, for events reconstructed in the central 0.6 cm radius region. The energy response was corrected only for the small attachment losses. The xenon X-ray
escape peak is clearly visible 30 keV below the full energy deposition peak. At 15 atm, a world record of 1 \% FWHM energy
resolution for 662 keV Gamma-rays was obtained. If scaled according to a $\frac{1}{\sqrt{E}}$ dependency, this resolution extrapolates to 0.52 \% at $Q_{\beta\beta}$, well below the initial target of the NEXT experiment (1~\%). 

Figure~\ref{fig:NEXTDBDM_EnRes_30_10bar} shows the spectrum for fluorescence X-rays pulses that were well separated in time from the main deposition and were reconstructed in the central 1.5 cm radius region. For the K$_\alpha$ peak, with a nominal energy of 29.6 keV~\cite{k_alpha}, an energy resolution of 5 \% FWHM was achieved. The centroid of this peak, after energy calibration on the 662 keV full
energy peak (and considering linearity with a zero intercept), is at 29.1 keV, less 2 \% away from the nominal value. The energy resolution at these energies is good enough to clearly resolve separately the K$_\alpha$ and K$_\beta$ (nominally at 33.6 keV) lines of X-ray shell transitions in xenon.

\begin{figure}[h]
\centering
\includegraphics[width=.5\textwidth]{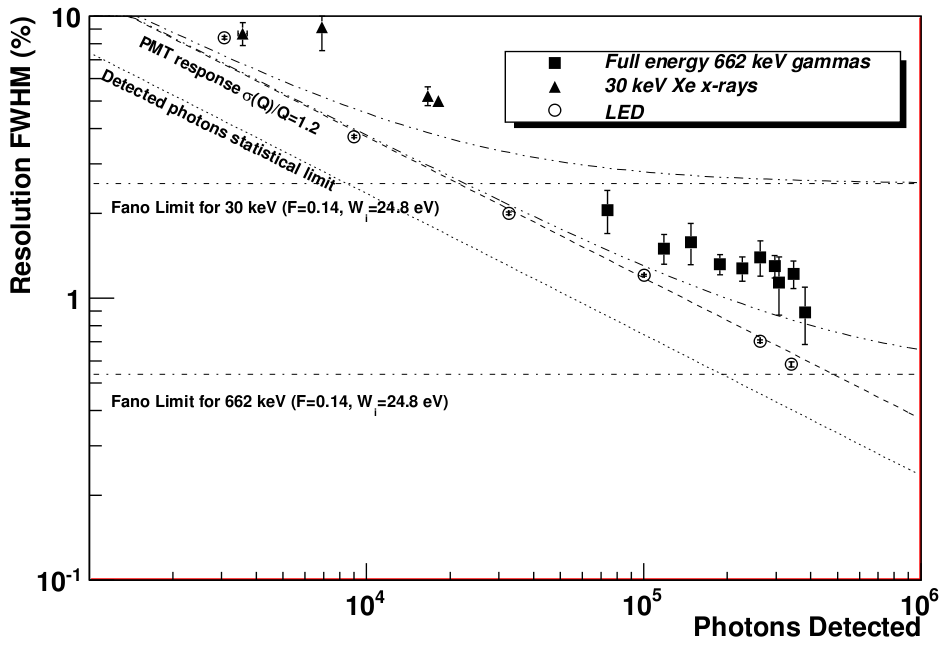}%
\caption{Understanding the energy response of the NEXT-DBDM: points show the measured energy resolutions for 662 keV Gammas, 30 keV xenon X-rays  and LED light pulses as a function of the number of detected photons. The expected resolution 
including the contributions from the intrinsic Fano factor, the statistical fluctuations in the number of 
detected photons and the PMT amplification fluctuations is shown for X-rays (dot dot dashed) and for 662 keV Gammas (dot dot dot dashed). The results referring to 662 keV depositions were obtained from 15.1 atm data runs while fluorescence X-ray results were  obtained from 10.1 atm runs.}
\label{fig:NEXTDBDM_EnRes_resume}
\end{figure}

The plot of Fig.~\ref{fig:NEXTDBDM_EnRes_resume} summarises our knowledge about the energy response in the NEXT-DBDM prototype. The energy resolution dependence on the amount of detected light in an electroluminescent TPC is described by four contributions. The first is the gas intrinsic energy resolution given by the Fano factor~\cite{fano} and the energy of the incoming particle, $F=0.14$ for our conditions. It is represented in the plot by the two horizontal lines, for 30 keV X-rays and 662 keV Gamma-rays respectively. The process of EL amplification is a nearly noiseless process and thus does not contribute significantly to the energy resolution degradation~\cite{EL_coliveira}. The conversion of photons into photoelectrons in the PMTs photocathode follow a Poisson distribution represented by the lower diagonal line. The upper diagonal line shows the expected response when including the variance in the PMT response to single photoelectrons, which we determined previously. The data points taken with a pulsed LED (circles) follow that trend. 

Summing all the contributions one gets the two curved lines, respectively for 30 and 662 keV depositions. The data points follow the expected tendency but are larger than the expected form (20-30 \% in the case of 662 keV Gammas) possibly due to the $\left(x,y\right)$ response non-uniformity. It is expected that a dense tracking array as that recently installed in the prototype, which preliminary results are presented in Section~\ref{subsec:tracking}, will allow one to map that non-uniformity and correct the detector response, further enhancing the energy resolution.

\subsection{Tracking}
\label{subsec:tracking}

\begin{figure}[ht]
\centering
\subfigure[Picture of the 8 $\times$ 8 array of SiPMs supported by 4 Dice-boards, placed behind the anode mesh. For a scale reference, the inner diagonal of the hexagonal frame is 14.4 cm long.]{
   \includegraphics[height=0.28\textwidth] {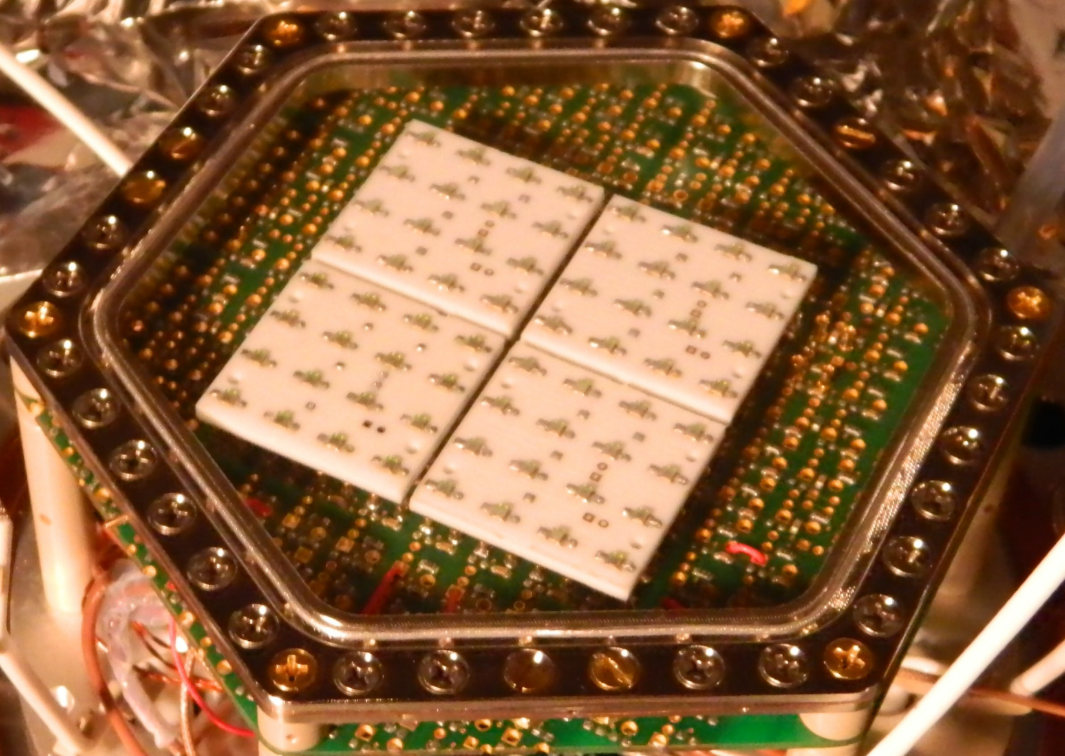}
   \label{fig:NEXTDBDM_SiPMs}
 }
 ~
 \subfigure[Picture of the low power electronics SiPM readout. A front-end card (at the foreground), used for signal amplification and shaping, connects to 4 SiPMs at once. Four of these are connected together to a multiplexing board (in the left of the picture).]{
   \includegraphics[height=0.28\textwidth] {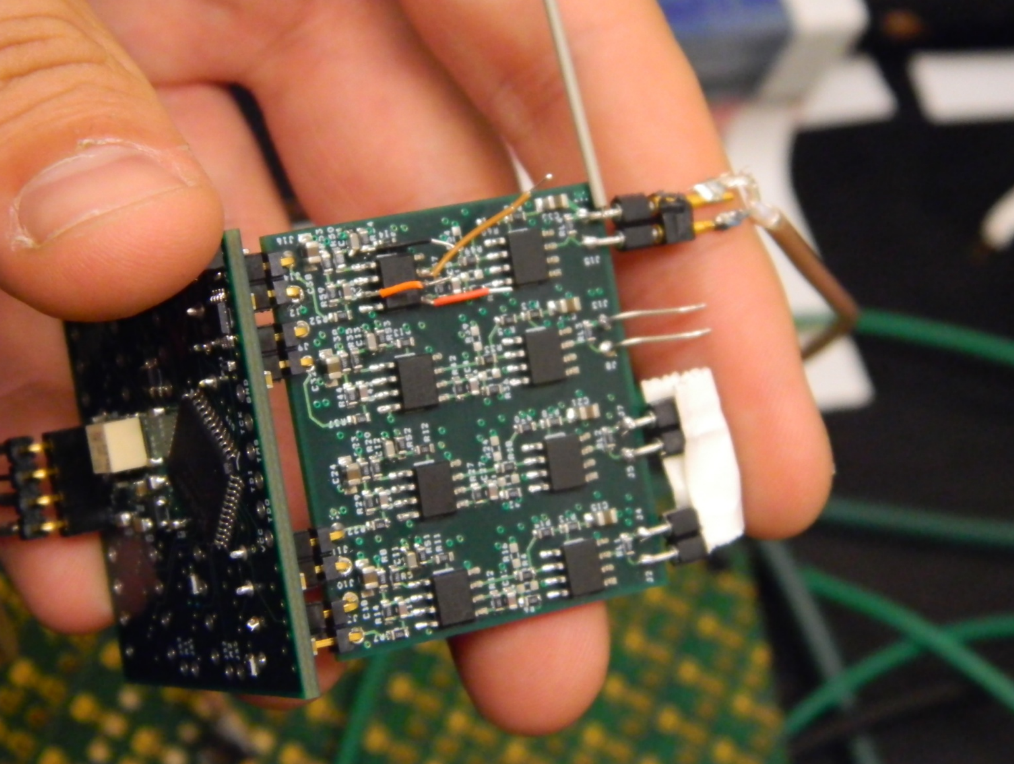}
   \label{fig:NEXTDBDM-SIPMs_MultiPlex}
 }
\label{fig:NEXTDBDM_tracking_array}
\caption{Tracking system of NEXT-DBDM.}
\end{figure}

In this Section, we report tracking results achieved after installing 64 Hamamatsu S10362-11-025P SiPMs, placed 2 mm behind the anode. These devices have 1 mm$^2$ active area and were arranged in a 8 $\times$ 8 square array with a pitch of 1 cm, as shown in Fig.~\ref{fig:NEXTDBDM_SiPMs}. The SiPMs have been coated with TPB so they can be sensitive to the 172 nm light from xenon EL~\cite{SiPMs_coating}.  A set of 4 Dice-boards provide support for the SiPMS as well as connections for high voltage supply and signal readout. Figure~\ref{fig:NEXTDBDM-SIPMs_MultiPlex} shows the low power electronics used for signal amplification, shaping and multiplexing of the SiPMs waveforms. These waveforms (not calibrated for the results presented here) are sliced in 30 keV energy equivalent sections and used to perform likelihood fits in order to determine the position of such partial depositions. 
 
\begin{figure}[ht]
\centering
 \subfigure[2D image of the track of an externally triggered muon. Circular points represent the S$_2$ intensities detected in each SiPM, according to the gray scale. The square points represent the reconstructed $\left(x,y\right)$ positions of 30 keV point-like depositions.]{
   \includegraphics[height=0.39\textwidth] {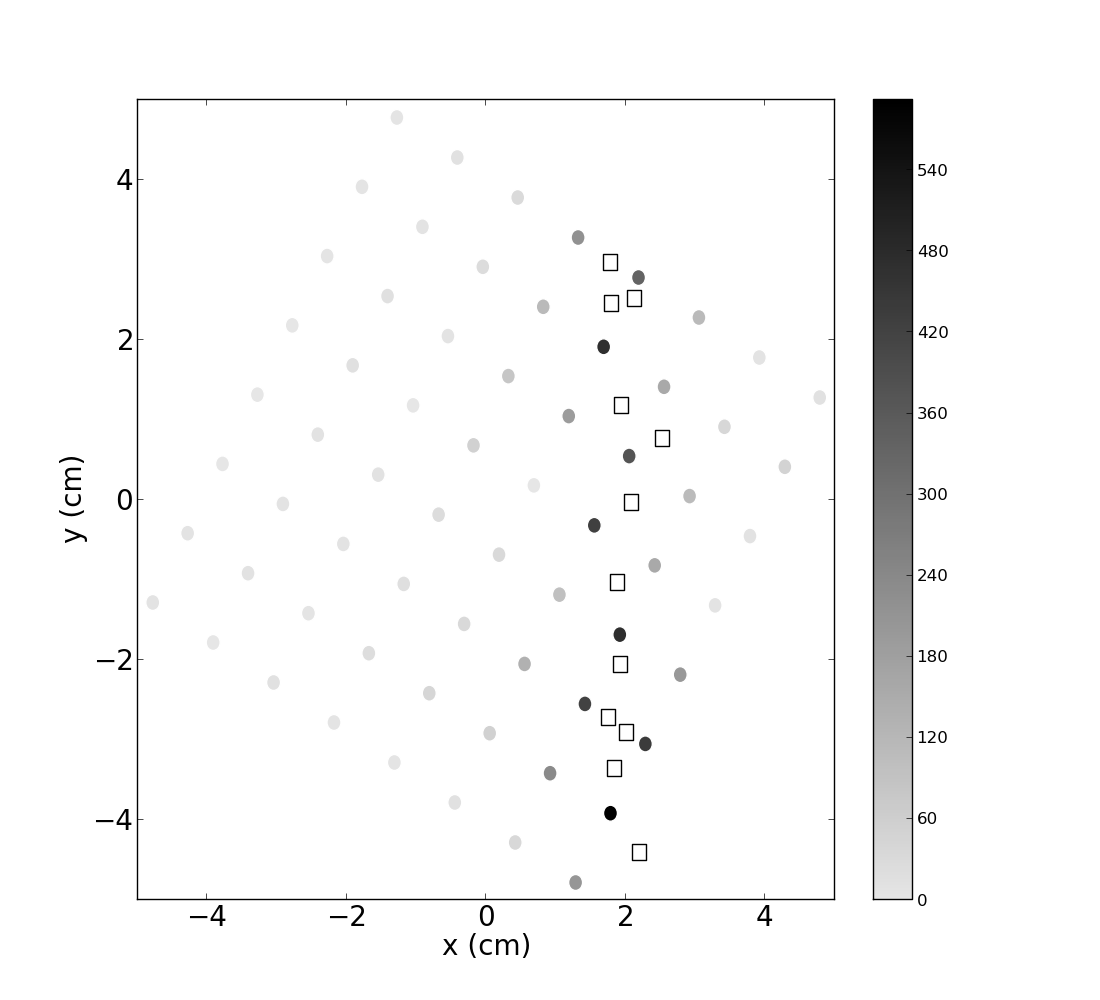}
   \label{fig:NEXTDBDM_muon}
 }
 ~
\subfigure[2D image of the deposition of a Gamma-ray. The main deposition at the center of chamber is visible along with a well separated fluorescence X-ray deposited in the up right edge of the SiPM array.
]{
   \includegraphics[height=0.39\textwidth] {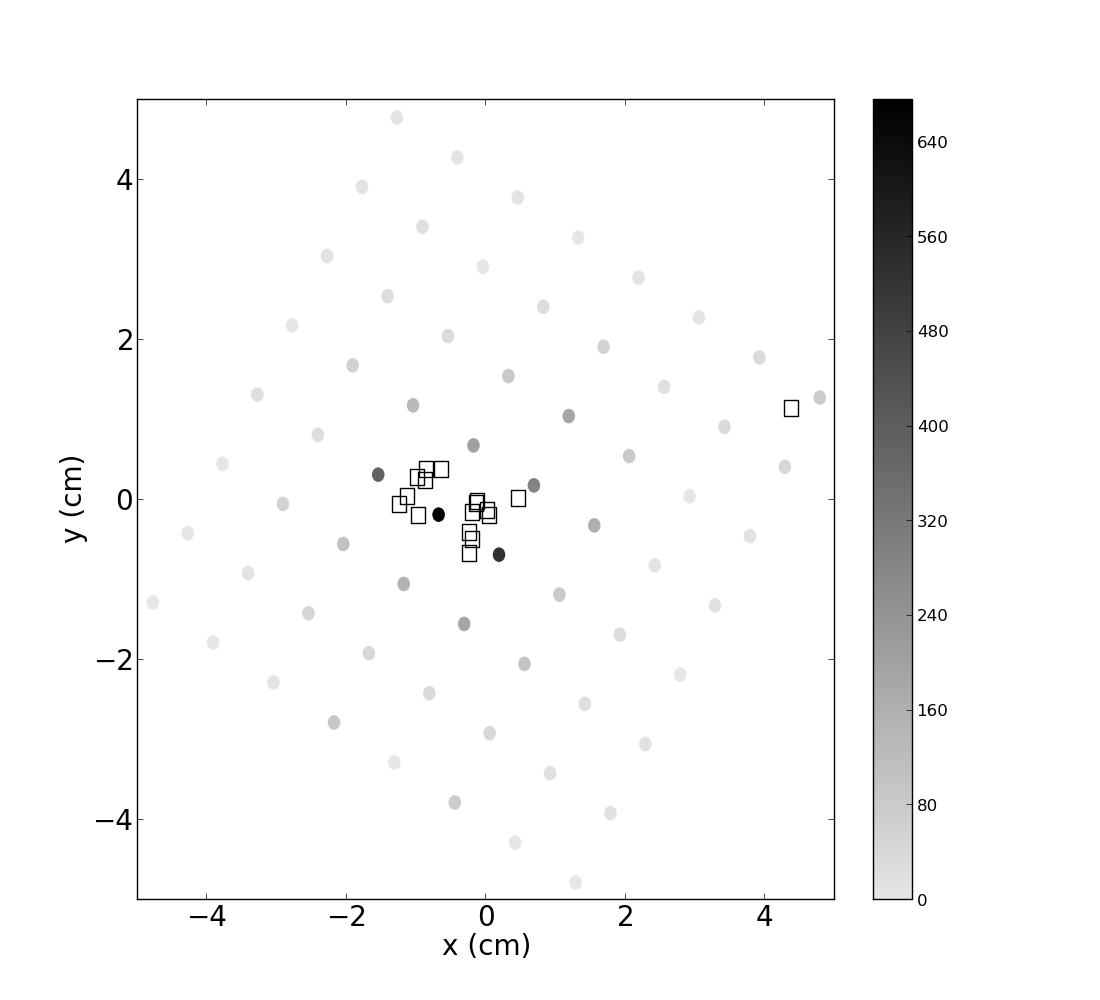}
   \label{fig:NEXTDBDM_gamma}
 }
\label{fig:NEXTDBDM_tracking}
\caption{Tracking with NEXT-DBDM.}
\end{figure}

Figure~\ref{fig:NEXTDBDM_muon} shows the relative S$_2$ intensities detected in each SiPM for an externally tagged cosmic ray muon traversing the chamber. The squares are the reconstructed $\left(x,y\right)$ positions of each section of the sliced data and show the expected straight muon trajectory. The estimated position RMS error is estimated to be 2.1 mm. 

In Fig.~\ref{fig:NEXTDBDM_gamma}, a similar plot is shown but for the case of a 662 keV Gamma-ray event. The main deposition in the center of the chamber is clearly visible. It is about 3 cm long and affected by the expected multiple scattering that makes the track curled. Around the position $\left(4,1.5\right)$ cm, it is also visible a well separated X-ray, about 5 cm away from the main deposition. 

These results, which can still be improved since they are preliminary, were obtained without calibration and with no optimized analysis, show that  tracking with SiPMs works in a gaseous electroluminescent TPC, enabling topological identification of the $0\nu\beta\beta$ decay.

\section{Other developments}
The drift velocity and longitudinal diffusion of electrons in pure xenon has been assessed in NEXT-DEMO using alpha particles, which produce point-like depositions and show a  strong anti-correlation between the ionization and the scintillation signals. Details about this work may be found in reference~\cite{alphas}. Using NEXT-DBDM and 30 keV fluorescence X-rays (also point-like) we measured the electron drift parameters in mixtures of xenon and CF$_4$, achieving good consistency with NEXT-DEMO and Monte Carlo results. Measurements of Xe+CH$_4$ are also planned and will be reported in a future publication.

As a parallel R\&D activity, we have been exploring the possibility of using MicroMEGAS, a Micropattern Gaseous Detector operating in avalanche mode, as a radiopure and low out-gassing device for both energy measurement and tracking. We were able to reconstruct Gamma-ray interaction tracks consistent in length with the expectations for energies between 14 and 122 keV when using a mixture of xenon and tri-methyl-amine (TMA) at 1 bar. The obtained energy resolutions extrapolate to 2.96 \% FWHM when scaled to $Q_{\beta\beta}$.

\section{Conclusions}

In the last 3 years the NEXT collaboration carried out a R\&D program with the aim of demonstrating the NEXT detector concept and gaining experience
that would facilitate the design, construction and operation of the NEXT-100 detector. Two prototypes, NEXT-DEMO and NEXT-DBDM, were constructed and operated, with complementary goals. 

NEXT-DEMO has shown the advantages of coating the TPC walls with TPB with a 3.0$\times$ improvement in S$_1$ and S$_2$ light collection. Outstanding energy resolution of 1.0 \% FWHM for 662 keV Gamma-rays was achieved with NEXT-DBDM for the range of xenon pressures that will be used in NEXT-100. This energy resolution extrapolates to 0.52~\% at $Q_{\beta\beta}$, well below the initial 1 \% goal of NEXT. Tracking using SiPMs for measuring the EL signal works and will allow further improvements in the energy resolution after correction of the S$_2$ position dependencies.

These promising results represent a milestone towards the success of NEXT-100 and prove the working concept of such a detector.

\ack

This work was supported by the Director, Office of
Science, Office of Basic Energy Sciences, of the U.S. Department of Energy under Contract No. DE-AC02-05CH11231. This work used resources of the National Energy Research Scientific Computing Center (NERSC). This work was also supported by the Ministerio de Econom\'{i}a y Competitividad of Spain under grants CONSOLIDER-Ingenio 2010 CSD2008-0037 (CUP) and FPA2009-13697-C04-04.

\end{document}